\documentclass[letter,twocolumn]{jpsj3}
\usepackage{bm,amssymb,amsmath}	
\usepackage{times}
\usepackage{color}

\title{Local Correlation Effects on the $s_{\pm}$- and $s_{++}$-wave Superconductivities\\ Mediated by Magnetic and Orbital Fluctuations\\ in the five-orbital Hubbard Model for Iron Pnictides}

\author{Jun Ishizuka$^1$, Takemi Yamada$^1$, Yuki Yanagi$^2$, Yoshiaki \=Ono$^1$ 
}
\inst{1. Department of Physics, Niigata University, Ikarashi, Niigata 950-2181, Japan \\
2. Department of Physics, Faculty of Science and Technology, Tokyo University of Science, Noda 278-8510, Japan 
}
\abst{
We investigate the electronic state and the superconductivity in the 5-orbital Hubbard model for iron pnictides by using the dynamical mean-field theory in conjunction with the Eliashberg equation. 
The renormalization factor exhibits significant orbital dependence resulting in the large change in the band dispersion as observed in recent ARPES experiments. 
The critical interactions towards the magnetic, orbital and superconducting instabilities are suppressed as compared with those from the random phase approximation (RPA) due to local correlation effects. 
Remarkably, the $s_{++}$-pairing phase due to the orbital fluctuation is largely expanded relative to the RPA result, while the $s_{\pm}$-pairing phase due to the magnetic fluctuation is reduced. 
}

\kword{dynamical mean-field theory, iron-based superconductor, spin fluctuation, orbital fluctuation, 5-orbital Hubbard model}
\begin{document}
\maketitle

The discovery of the iron-based superconductors\cite{kamihara} has triggered an intense research effort to investigate their electronic state and superconducting mechanism. Most of the phase diagrams exhibit the tetragonal-orthorhombic structural transition and the stripe-type antiferromagnetic (AFM) transition\cite{kamihara,jhonston}. The AFM fluctuation is enhanced towards the AFM transition\cite{ning}, while the ferro-orbital (FO) fluctuation responsible for the softening of the elastic constant $C_{66}$\cite{yoshizawa_1,goto} is enhanced towards the structural transition. Correspondingly, two distinct $s$-wave pairings: the $s_{\pm}$-wave with sign change of the order parameter between the hole and the electron Fermi surfaces (FSs) mediated by the AFM fluctuation \cite{mazin,kuroki} and the $s_{++}$-wave without the sign change mediated by the FO fluctuation\cite{yanagi_3,yanagi_4,yanagi_2} and by the antiferro-orbital (AFO) fluctuation\cite{kontani} which is also responsible for the softening of $C_{66}$ through the two-orbiton process\cite{kontani_2} were proposed. Despite the numerous efforts, the pairing state together with the mechanism of the superconductivity is still controversial.

As the details of the electronic band structure are crucial for the pairing state and mechanism, the theoretical studies have employed the realistic multi-orbital models\cite{kuroki,yanagi_3,yanagi_4,yanagi_2,kontani,kontani_2} where the tight-binding parameters are determined so as to reproduce the first-principles band structures which had been found to agree with the angle-resolved photoemission spectroscopy (ARPES) by reducing the band width by a factor of $2\sim 3$\cite{Nishi}. However, recent high-resolution ARPES measurements for Ba$_{0.6}$K$_{0.4}$Fe$_2$As$_2$\cite{ding_2} revealed significant band (or orbital) dependence of the mass enhancement from 1.3 to 9.0. 
More recently, some evidences for an orbital-selective Mott transition (OSMT) in K$_x$Fe$_{2-y}$Se$_2$\cite{yi_2}, where the renormalization factor $Z$ for $d_{xy}$ orbital becomes zero while $Z$ for the other orbitals are finite, and for the heavy fermion behavior in KFe$_{2}$As$_2$\cite{Hardy}, where the system is near the OSMT, were observed. In these cases, we need to investigate the superconductivity on the basis of the strongly correlated electronic states in the presence of the large orbital dependence of $Z$.

In this letter, we investigate the 5-orbital Hubbard model\cite{kuroki} for iron pnictides by using the dynamical mean-field theory (DMFT)\cite{georges} which is exact in infinite dimensions ($d=\infty$) where the self-energy becomes local and enables us to sufficiently take into account the local correlation effects including the strong correlation regime where $Z$ largely depends on the orbital\cite{yin} and the OSMT is realized\cite{koga}. 
To examine the superconductivity, we solve the Eliashberg equation in which the effective pairing interaction and the renormalized single-particle Green's function are calculated on the basis of the DMFT. 
In particular, we focus our attention on the local correlation effects on the possible pairing states, the magnetic fluctuation mediated $s_{\pm}$-wave and the orbital fluctuation mediated $s_{++}$-wave, beyond the random phase approximation (RPA) which was extensively developed for iron pnictides in the previous works\cite{kuroki,yanagi_3,yanagi_4,yanagi_2,kontani}.


The 5-orbital Hubbard model consists of the Fe $3d$-orbitals and is given by the Hamiltonian\cite{kuroki}, $\hat{H}=\hat{H}_0+\hat{H}_{\rm int}$, 
where the kinetic part $\hat{H}_0$ is determined so as to reproduce the first-principles band structure for LaFeAsO and the Coulomb interaction part $\hat{H}_{\rm int}$ includes the multi-orbital interaction on a Fe site: the intra- and inter-orbital direct terms $U$ and $U'$, Hund's rule coupling $J$ and the pair transfer $J'$. 
In this paper, we set the $x$-$y$ axes parallel to the nearest Fe-Fe bonds.

To solve the model, we use the DMFT\cite{georges} in which the lattice model is mapped onto an impurity Anderson model embedded in an effective medium which is determined so as to satisfy the self-consistency condition: 
$
{\hat G}(i\varepsilon_m)=(1/N)\sum_{\bm{k}}{\hat G}({\bm k},i\varepsilon_m)
$
with the wave vector $\bm{k}$ and the Matsubara frequency $\varepsilon_m=(2m+1)\pi T$, where ${\hat G}(i\varepsilon_m)$ and ${\hat G}({\bm k},i\varepsilon_m)$ are the 5$\times$5 matrix representations of the local (impurity) Green's function and the lattice Green's function, respectively, which are explicitly given by
\begin{align}
{\hat G}(i\varepsilon_m)&=\left[{\cal \hat{G}}^{-1}(i\varepsilon_m) 
-{\hat \Sigma}(i\varepsilon_m)\right]^{-1},
\label{eq:G_loc}\\
{\hat G}({\bm k},i\varepsilon_m)&=\left[(i\varepsilon_m+\mu)-{\hat H}_0({\bm k})
-{\hat \Sigma}(i\varepsilon_m)\right]^{-1}, 
\label{eq:G}
\end{align} 
where ${\hat \Sigma}(i\varepsilon_m)$ is the $5\times 5$ matrix representation of the impurity (local) self-energy and ${\cal\hat{G}}(i\varepsilon_m)$ is that of the bare impurity Green's function describing the effective medium. 
Within the DMFT, the spin (charge-orbital) susceptibility is given in the $25\times 25$ matrix representation as 
\begin{eqnarray}
\hat{\chi}_{s(c)}(q)=
\left[1 -(+)\hat{\chi}_0(q)\hat{\Gamma}_{s(c)}(i\omega_n)\right]^{-1} 
\hat{\chi}_0(q)
\label{eq:chi}
\end{eqnarray}
with 
$
\hat{\chi}_0(q)=-(T/N)\sum_{k}\hat{G}(k+q)\hat{G}(k), 
$
where $k=(\bm{k},i\varepsilon_m)$, $q=(\bm{q},i\omega_n)$ and $\omega_n=2n\pi T$.  In eq. (\ref{eq:chi}), $\hat{\Gamma}_{s(c)}(i\omega_n)$ is the local irreducible spin (charge-orbital) vertex in which only the external frequency ($\omega_n$) dependence is considered as a simplified approximation\cite{park} and is explicitly given by
\begin{eqnarray}
\hat{\Gamma}_{s(c)}(i\omega_n)=-(+)\left[\hat{\chi}_{s(c)}^{-1}(i\omega_n)-\hat{\chi}_0^{-1}(i\omega_n)\right]
\label{eq:gamma}
\end{eqnarray} 
with
$
\hat{\chi}_0(i\omega_n)=-T\sum_{\varepsilon_m}
\hat{G}(i\varepsilon_m+i\omega_n)\hat{G}(i\varepsilon_m), 
$
where $\hat{\chi}_{s(c)}(i\omega_n)$ is the local spin (charge-orbital) susceptibility. 
When the largest eigenvalue $\alpha_s$ ($\alpha_c$) of  $(-)\hat{\chi}_0(q)\hat{\Gamma}_{s(c)}(i\omega_n)$ in eq. (\ref{eq:chi})  for a wave vector $\bm{q}$ with $i\omega_n=0$ reaches unity, the instability towards the magnetic (charge-orbital) order with the corresponding $\bm{q}$ takes place.

To examine the superconductivity mediated by the magnetic and charge-orbital fluctuations which are extremely enhanced towards the corresponding orders mentioned above, we write the effective pairing interaction for the spin-singlet state using the spin (charge-orbital) susceptibility and vertex given in eqs. (\ref{eq:chi}) and (\ref{eq:gamma}) obtained within the DMFT in the $25\times 25$ matrix representation as
\begin{align}
\hat{V}(q)
&=\frac{3}{2}\hat{\Gamma}_{s}(i\omega_n)\hat{\chi}_{s}(q)\hat{\Gamma}_{s}(i\omega_n)
-\frac{1}{2}\hat{\Gamma}_{c}(i\omega_n)\hat{\chi}_{c}(q)\hat{\Gamma}_{c}(i\omega_n) \nonumber \\
&+\frac{1}{2}\left(\hat{\Gamma}_{s}^{(0)}+\hat{\Gamma}_{c}^{(0)}\right)
\label{eq:pair}
\end{align}
with the bare spin (charge-orbital) vertex: 
$[\Gamma_{s(c)}^{(0)}]_{\ell\ell\ell\ell}=U(U)$, 
$[\Gamma_{s(c)}^{(0)}]_{\ell\ell'\ell\ell'}=U'(-U'+2J)$, 
$[\Gamma_{s(c)}^{(0)}]_{\ell\ell\ell'\ell'}=J(2U'-J)$ and 
$[\Gamma_{s(c)}^{(0)}]_{\ell\ell'\ell'\ell}=J'(J')$, 
where $\ell'\neq \ell$ and the other matrix elements are 0. 
Substituting the effective pairing interaction eq. (\ref{eq:pair}) and the lattice Green's function eq. (\ref{eq:G}) into the linearized Eliashberg equation: 
\begin{align}
\lambda \Delta_{ll'}(k)&=-\frac{T}{N}\sum_{k'}
\sum_{l_1l_2l_3l_4}V_{ll_1,l_2l'}(k-k') \nonumber \\
&\times G_{l_3l_1}(-k')\Delta_{l_3l_4}(k') G_{l_4l_2}(k'), 
\label{gapeq}
\end{align}
we obtain the gap function $\Delta_{ll'}(k)$ with the eigenvalue $\lambda$ 
which becomes unity at the superconducting transition temperature $T=T_c$.  
In eq. (\ref{gapeq}), $\Delta_{ll'}(k)$ includes the $1/d$ corrections yielding the ${\bm k}$ dependence of the gap function responsible for the anisotropic superconductivity which is not obtained within the zeroth order of $1/d$\cite{georges}. 
If we replace $\hat{\Gamma}_{s(c)}$ with $\hat{\Gamma}_{s(c)}^{(0)}$ and neglect $\hat{\Sigma}$, eq. (\ref{eq:pair}) yields the RPA result of $\hat{V}(q)$\cite{kuroki,yanagi_3,yanagi_4,yanagi_2,kontani}. 
Therefore, eq. (\ref{gapeq}) with eqs. (\ref{eq:G}) and (\ref{eq:pair}) is a straightforward extension of the RPA to include the vertex and the self-energy corrections within the DMFT without any double counting. 

In the actual calculations with the DMFT, 
we solve the effective 5-orbital impurity Anderson model, where the Coulomb interaction at the impurity site is given by the same form as $\hat{H}_{\rm int}$ with a site $i$ and the kinetic energy responsible for ${\cal \hat{G}}$ in eq. (\ref{eq:G_loc}) is determined so as to satisfy the self-consistency condition as possible, by using the exact diagonalization (ED) method for a finite-size cluster to obtain the local quantities such as $\hat{\Sigma}$ and $\hat{\chi}_{s(c)}$. Since the multi-orbital system requires rather CPU-time and memory consuming calculations, we employ the clusters with the site number $N_s=4$ within a restricted Hilbert space\cite{yamada_2}. We have also performed preliminary calculations with $N_s=2$\cite{ishizuka} and have confirmed that the results with $N_s=4$ are qualitatively consistent with those with $N_s=2$ and quantitatively improved especially for the intermediate interaction regime as previously observed in the DMFT+ED approaches for the multi-band and multi-orbital models\cite{ono_1,ono_2,ono_3,liebsch_3}. 
In fact, the DMFT results from the ED with $N_s=4$ are quantitatively in good agreement with the precise results from the numerical renormalization group\cite{pruschke} for the 2-orbital Hubbard model and those from the continuous-time quantum Monte Carlo\cite{werner} for the 3-orbital Hubbard model\cite{liebsch_3}. As for the 5-orbital Hubbard model, the ED results with $N_s=3$ are found to agree with those with $N_s=2$\cite{liebsch_3}. Therefore, we expect that the ED calculations with $N_s=4$ yield quantitatively reliable results also for the present 5-orbital Hubbard model.
All calculations are performed at $T=0.02{\rm eV}$ for the electron number $n=6.0$ corresponding to the non-doped case. We use $32\times 32$ ${\bm k}$-point meshes and 1024 Matsubara frequencies in the numerical calculations with the fast Fourier transformation. Here and hereafter, we measure the energy in units of eV.


In the previous RPA study\cite{yanagi_3}, it was found that the $s_{\pm}$-pairing is mediated by the magnetic fluctuation near the AFM order for $U>U'$, while the $s_{++}$-pairing is mediated by the orbital fluctuation near the FO order for $U<U'$, where the superconductivity is investigated in the wide parameter space by treating $U$, $U'$, $J$ and $J'$ as independent parameters apart from the condition satisfied in the isolated atom: $U=U'+2J$ and $J=J'$. 
Correspondingly, we consider the two specific cases with $U>U'$ and $U<U'$ to elucidate the correlation effects beyond the RPA on the magnetic and orbital orders and the those fluctuations mediated superconductivity.

\begin{figure}[t]
\begin{center}
\includegraphics[width=60mm]{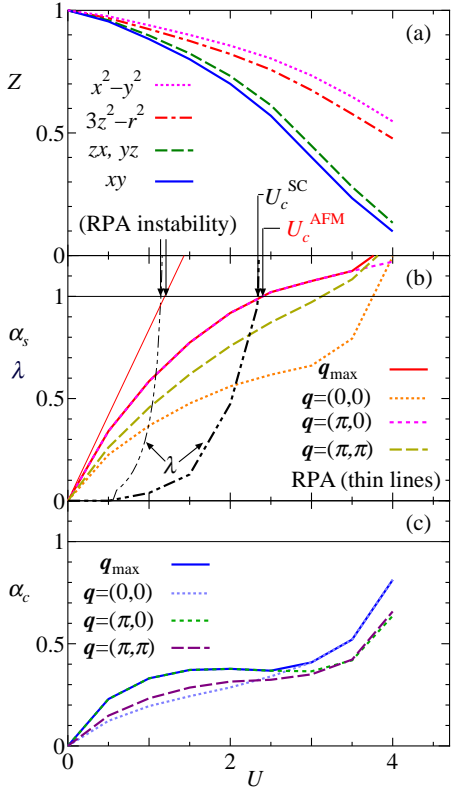}
\caption{(Color online) 
(a) The renormalization factor $Z_{\ell}$ with $\ell=d_{x^2-y^2}, d_{3z^2-r^2}, d_{zx}, d_{yz}$ and $d_{xy}$, (b) and (c)  the largest eigenvalues $\alpha_s$ and $\alpha_c$ for several ${\bm q}$ and $\lambda$ which reach unity towards the magnetic, charge-orbital and superconducting instabilities, respectively, as functions of $U$ with $U=U'+2J$, $J/U=0.1$ and $J=J'$ for $n=6.0$ and $T=0.02$. The RPA results of $\alpha_s$ for ${\bm q}_{\rm max}$ and $\lambda$ are also plotted by thin lines in (b). 
\label{fig_udep}}
\end{center}
\end{figure}

First, we consider the case with $U>U'$, where the magnetic fluctuation dominates over the orbital fluctuation. In Fig. \ref{fig_udep}, several physical quantities are plotted as functions of $U$ with $U=U'+2J$, $J/U=0.1$ and $J=J'$. 
Fig. \ref{fig_udep} (a) shows the renormalization factor defined by: 
$Z_{\ell}=\left[1-\frac{d\Sigma_{\ell}(\varepsilon)}{d(\varepsilon)}\bigl.\bigr|_{\varepsilon\rightarrow0}\right]^{-1}$ with orbital $\ell=d_{x^2-y^2}, d_{3z^2-r^2}, d_{zx}, d_{yz}$ and $d_{xy}$. When $U$ increases, all of $Z_{\ell}$ monotonically decrease with increasing the variance of $Z_{\ell}$. We find that $Z_{\ell}$ for $\ell =d_{xy}$ is the smallest for all $U$ and finally becomes zero at $U_c \sim 5$ while $Z_{\ell}$ for $\ell \ne d_{xy}$ are finite revealing the OSMT\cite{yamada_2}, as recently discussed in K$_x$Fe$_{2-y}$Se$_2$\cite{yi_2} and KFe$_{2}$As$_2$\cite{Hardy} where the ARPES experiments are well accounted for by the slave-spin mean-field\cite{yi_2,medici_1} and the slave-boson mean-field (Gutzwiller)\cite{Hardy} approximations yielding the OSMT with $Z_{d_{xy}}\to 0$. We note that, even in the intermediate correlation regime away from the OSMT, the large orbital dependence of $Z_{\ell}$ results in the significant change in the band dispersion\cite{yamada_2} which is consistent with the recent high-resolution ARPES measurements for Ba$_{0.6}$K$_{0.4}$Fe$_2$As$_2$\cite{ding_2}.

Figs. \ref{fig_udep} (b) and (c) show the $U$ dependence of the largest eigenvalues $\alpha_s$ and $\alpha_c$ for several wave vectors ${\bm q}$, where $\alpha_{s(c)}$ shows a maximum at ${\bm q}={\bm q}_{\rm max}$. 
When $U$ increases, both $\alpha_s$ and $\alpha_c$ increase with $\alpha_s>\alpha_c$ and $\alpha_s$ becomes unity at $U_c^{\rm AFM}\sim 2.40$ where the magnetic susceptibility with ${\bm q}\sim (\pi,0)$ corresponding to the stripe-type AFM diverges. The largest eigenvalue $\lambda$ of the Eliashberg equation (\ref{gapeq}) is also plotted in Fig. \ref{fig_udep} (b) and is found to increase with increasing $\alpha_s$ and finally reaches unity at $U_c^{\rm SC}\sim 2.34$ where the superconducting instability occurs. 
For comparison, we also plot the RPA results of $\alpha_s$ for ${\bm q}_{\rm max}$ and $\lambda$ in Fig. \ref{fig_udep} (b) and find that the critical interactions $U_c^{\rm AFM}$ and $U_c^{\rm SC}$ from the DMFT are about twice larger than those from the RPA\cite{kuroki} due to the correlation effects beyond the RPA and are consistent with the values of the effective Coulomb interactions derived from the downfolding scheme based on first-principles calculations\cite{miyake}.

\begin{figure}[t]
\begin{center}
\includegraphics[width=58mm,angle=-90]{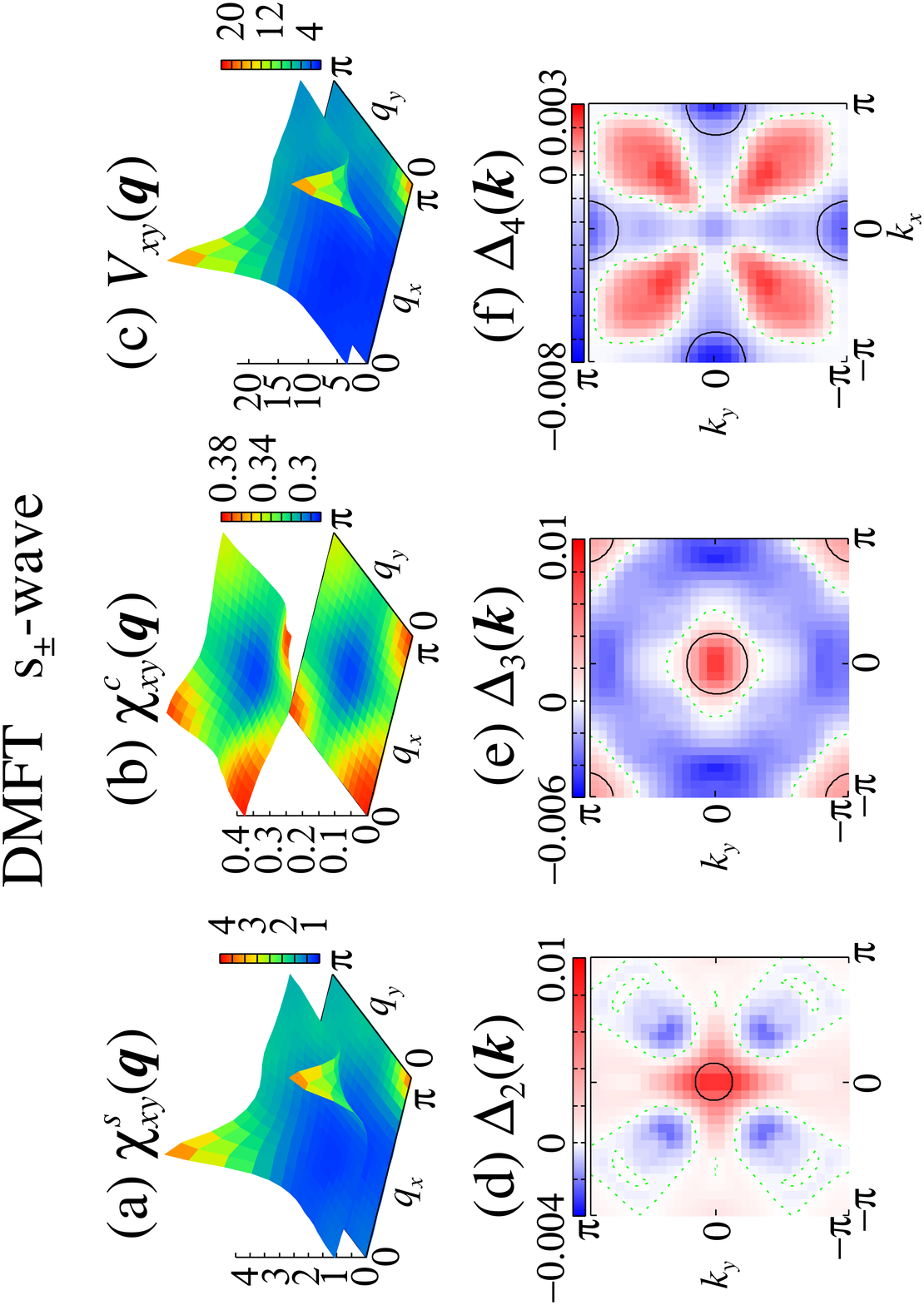}

\vspace{2mm}
\includegraphics[width=58mm,angle=-90]{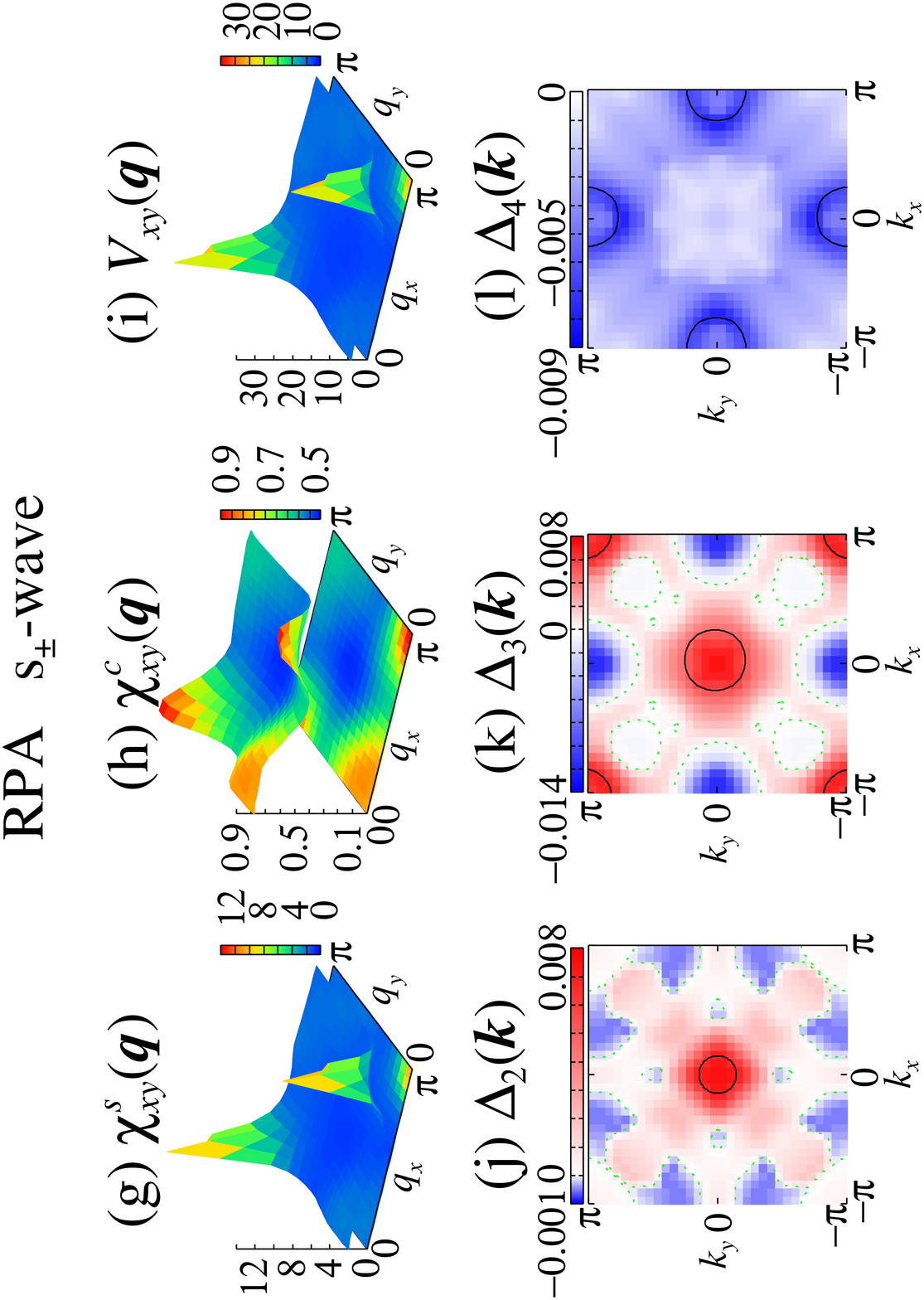}
\caption{(Color online) 
DMFT results for the $d_{xy}$ intra-orbital components of the spin susceptibility $\chi^{s}$ (a), the charge-orbital susceptibility $\chi^{c}$ (b) and the pairing interaction $V$ (c), and those for the band-diagonal components of the gap function $\Delta$  with the lowest Matsubara frequency $i\varepsilon_m=i\pi T$ for band 2 (d) and band 3 (e) (band 4 (f)) with the hole (electron) FSs (solid lines) for $U=2.28$, $U'=1.824$ and $J=J'=0.228$, where $\alpha_s=0.964$ for ${\bm q}_{\rm max}$. (g)-(l) The corresponding RPA results for $U=1.15$, $U'=0.92$ and $J=J'=0.115$, where $\alpha_s=0.964$ for ${\bm q}_{\rm max}$. 
\label{fig_gap1}}
\end{center}
\end{figure}

In Figs. \ref{fig_gap1} (a)-(f), we show the $d_{xy}$ intra-orbital components of the spin (charge-orbital) susceptibility $\chi^{s}$ ($\chi^{c}$) and the pairing interaction $V$, together with the band-diagonal components of the gap functions $\Delta$ with the lowest Matsubara frequency $i\varepsilon_m=i\pi T$ for $U=2.28$, $U'=1.824$ and $J=J'=0.228$. In this case, the enhanced spin susceptibility for ${\bm q}\sim (\pi,0)$, i. e., the stripe-type AFM fluctuation yields the large positive value of the effective pairing interaction $V$ for ${\bm q}\sim (\pi,0)$ resulting in the gap function with sign change between the electron and hole FSs, i. e.,  the $s_{\pm}$-wave state. Figs. \ref{fig_gap1} (g)-(l), we also show the corresponding RPA results for $U=1.15$, $U'=0.92$ and $J=J'=0.115$. As the ${\bm q}$ dependence of $\chi^{s}$ and $V$ from the DMFT becomes weak as compared to the RPA results due to the local correlation effects, the $s_{\pm}$-pairing phase is reduced relative to the RPA result as shown in Fig. \ref{fig_udep} (b).


\begin{figure}[t]
\begin{center}
\includegraphics[width=60mm]{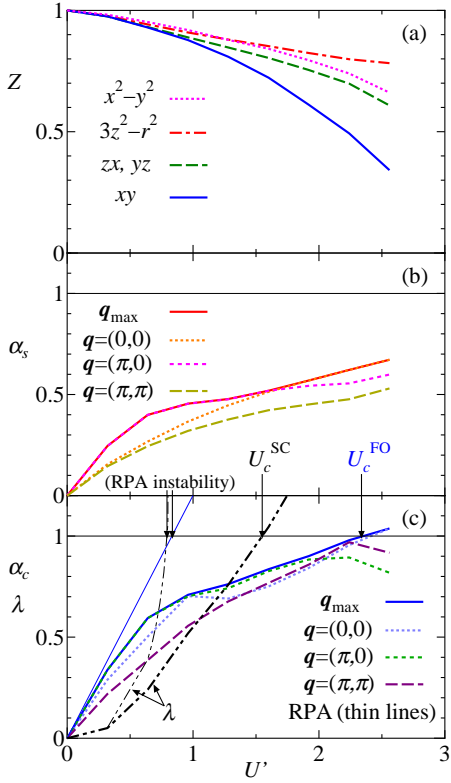}
\caption{(Color online)
(a) The renormalization factor $Z_{\ell}$ with $\ell=d_{x^2-y^2}, d_{3z^2-r^2}, d_{zx}, d_{yz}$ and $d_{xy}$, (b) and (c)  the largest eigenvalues $\alpha_s$ and $\alpha_c$ for several ${\bm q}$ and $\lambda$ which reach unity towards the magnetic, charge-orbital and superconducting instabilities, respectively, as functions of $U'$ with $U=0.25U'+2J$, $J/U=0.1$ and $J=J'$ for $n=6.0$ and $T=0.02$. The RPA results of $\alpha_c$ for ${\bm q}_{\rm max}$ and $\lambda$ are also plotted by thin lines in (c). 
\label{fig_u,dep}}
\end{center}
\end{figure}

Next, we consider the case with $U<U'$, where the orbital fluctuation dominates over the magnetic fluctuation. 
Figs. \ref{fig_u,dep} (a)-(c) show the renormalization factor $Z_{\ell}$ and the largest eigenvalues $\alpha_s$, $\alpha_c$ and $\lambda$ as functions of $U'$ with $U=0.25U'+2J$, $J/U=0.1$ and $J=J'$. 
When $U'$ increases, $Z_{\ell}$ for all $\ell$ monotonically decrease with keeping the smallest value for $\ell=d_{xy}$, similar to the case of Fig. \ref{fig_udep} (a). When $U'$ increases, both $\alpha_s$ and $\alpha_c$ increase with $\alpha_s<\alpha_c$ and $\alpha_c$ becomes unity at $U_c^{\rm FO}\sim 2.28$ where the orbital susceptibility with ${\bm q}\sim (0,0)$ corresponding to the FO diverges. 
We note that ${\bm q}_{\rm max}=(0,\pi/4)$ just below $U_c^{\rm FO}$ with $\alpha_c=0.98$ and ${\bm q}_{\rm max}=(0,0)$ just above $U_c^{\rm FO}$ with $\alpha_c=1.03$, while it is difficult to determine ${\bm q}_{\rm max}$ precisely at $U_c^{\rm FO}$ with $\alpha_c=1$ within the present numerical resolution as $\chi_c$ diverges almost simultaneously for ${\bm q}\sim (0,0)$  and then we call the FO in a broad sense.
With increasing $\alpha_c$, $\lambda$ increases and finally reaches unity at $U_c^{\rm SC}\sim 1.54$ where the superconducting instability occurs. 
For comparison, we also plot the RPA results of $\alpha_c$ for ${\bm q}_{\rm max}$ and $\lambda$ in Fig. \ref{fig_u,dep} (c), and find that $U_c^{\rm FO}$ and $U_c^{\rm SC}$ from the DMFT are larger than those from the RPA due to the correlation effects beyond the RPA. 
Remarkably, the DMFT result of the $s_{++}$-pairing phase with $U_c^{\rm SC}<U<U_c^{\rm FO}$ is largely expanded as compared to the RPA result, in contrast to the case with the $s_{\pm}$-pairing phase which is reduced (see Fig. \ref{fig_udep} (b)).

\begin{figure}[t]
\begin{center}
\includegraphics[width=58mm,angle=-90]{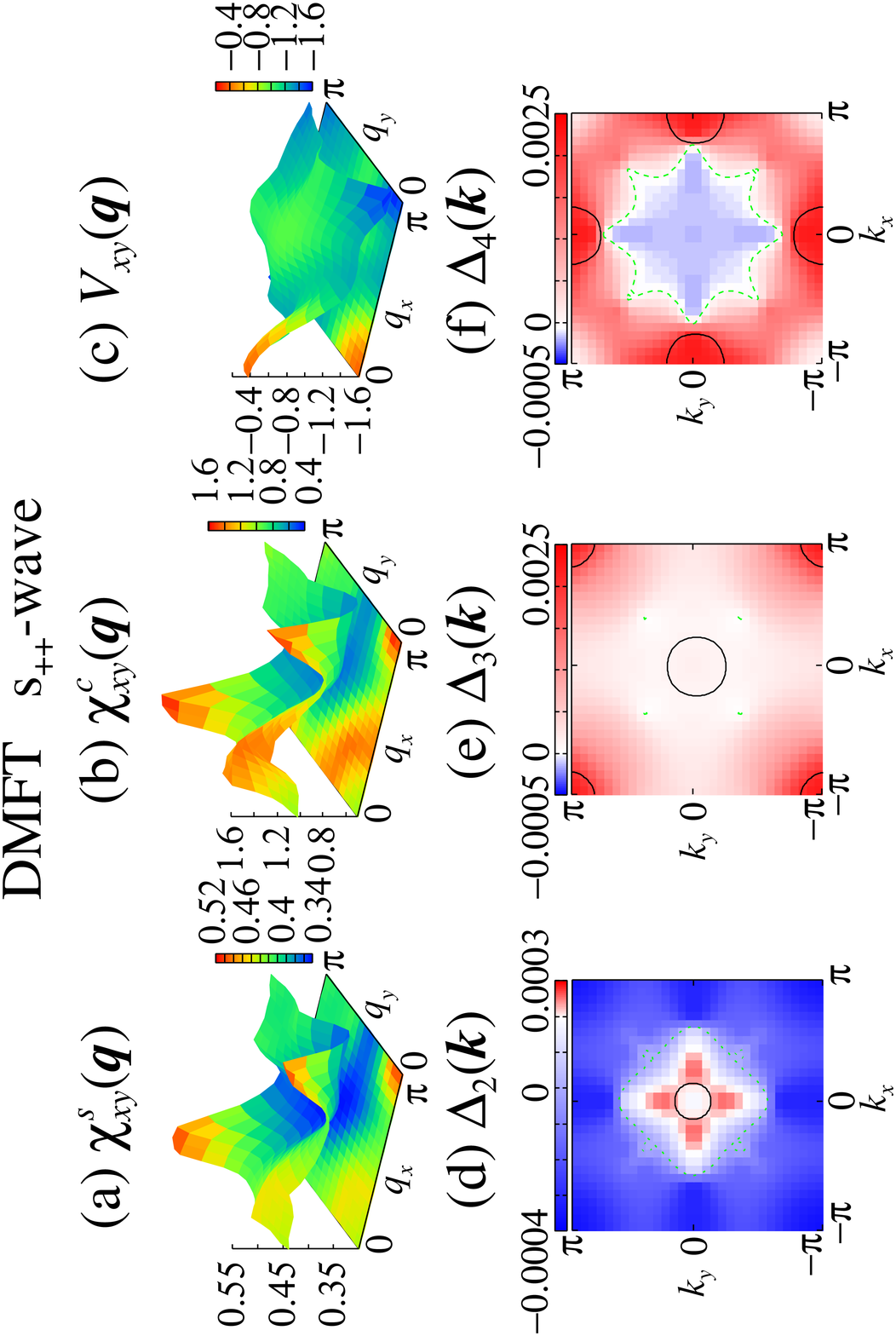}

\vspace{2mm}
\includegraphics[width=58mm,angle=-90]{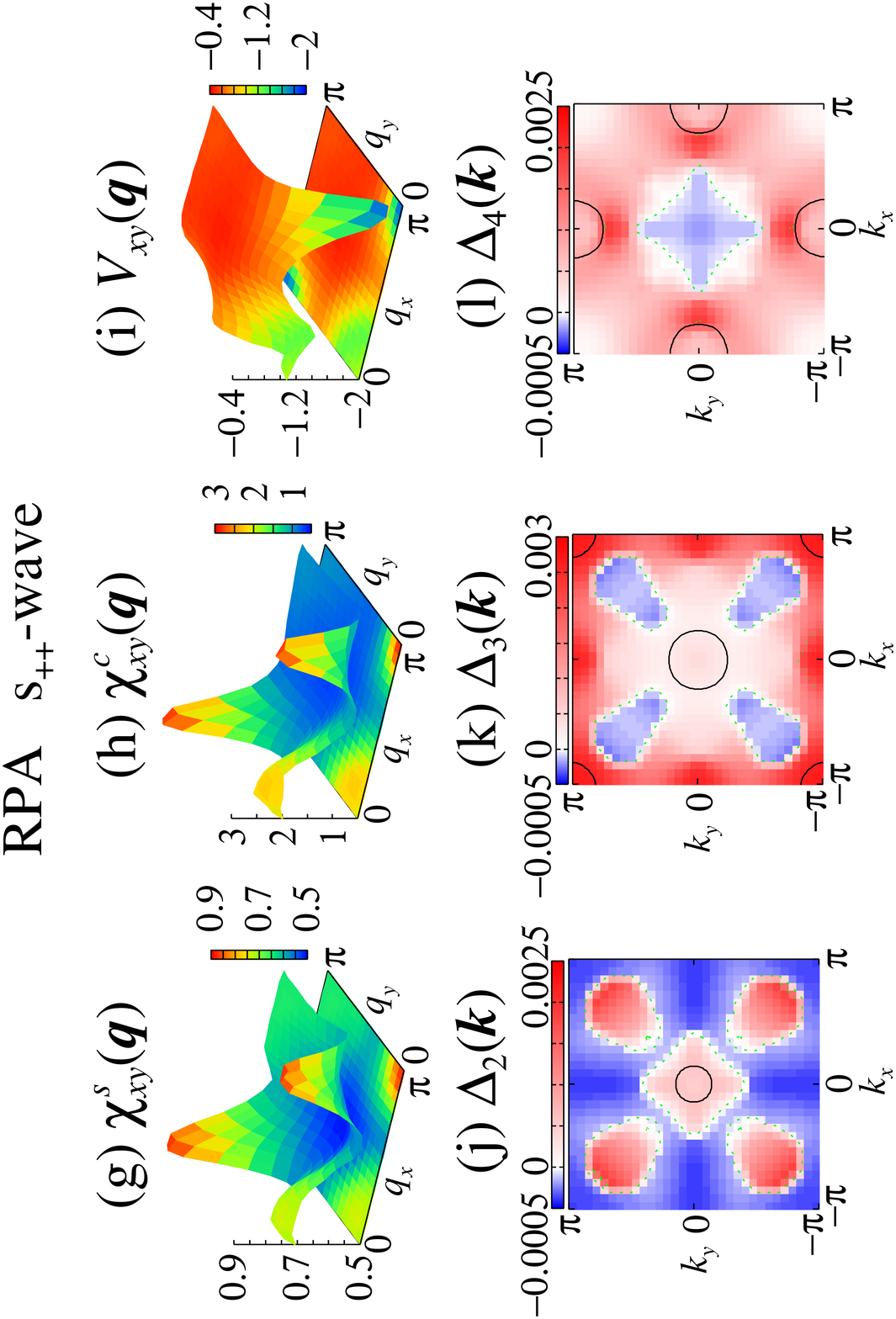}
\caption{(Color online) 
DMFT results for the $d_{xy}$ intra-orbital components of the spin susceptibility $\chi^{s}$ (a), the charge-orbital susceptibility $\chi^{c}$ (b) and the pairing interaction $V$ (c), and those for the band-diagonal components of the gap function $\Delta$ with $i\varepsilon_m=i\pi T$ for band 2 (d) and band 3 (e) (band 4 (f)) with the hole (electron) FSs (solid lines) for $U=0.4$, $U'=1.28$ and $J=J'=0.04$, where $\alpha_c=0.76$ for ${\bm q}_{\rm max}$. (g)-(l) The corresponding RPA results for $U=0.25$, $U'=0.8$ and $J=J'=0.025$, where $\alpha_c=0.76$ for ${\bm q}_{\rm max}$. 
\label{fig_gap2}}
\end{center}
\end{figure}

In Figs. \ref{fig_gap2} (a)-(f), we show the same physical quantities as in Figs. \ref{fig_gap1} (a)-(f) for $U=0.4$, $U'=1.28$ and $J=J'=0.04$. In this case, the enhanced orbital susceptibility in the whole ${\bm q}$ space yields the negative value of the effective pairing interaction $V$ for all ${\bm q}$ resulting in the gap function without sign change, i. e.,  the $s_{++}$-wave state. In Figs. \ref{fig_gap2} (g)-(l), we also show the corresponding RPA results for $U=0.25$, $U'=0.8$ and $J=J'=0.025$. As the ${\bm q}$ dependence of $\chi^{c}$ from the DMFT becomes weak as compared to the RPA result due to the local correlation effects, the local (${\bm q}$-averaged) component of the pairing attraction $|V|$  becomes considerably larger than the RPA result for the same value of $\alpha_c$ for ${\bm q}_{\rm max}$ resulting in the remarkable enhancement of the $s_{++}$-pairing phase which is expanded far away from the FO critical interaction $U_c^{\rm FO}$ ($\alpha_c=0.82$ for $U_c^{\rm SC}$) in contrast to the RPA result ($\alpha_c=0.95$ for $U_c^{\rm SC}$) as shown in Fig. \ref{fig_u,dep} (c).


In summary, we have investigated the electronic state and the superconductivity in the 5-orbital Hubbard model for iron pnictides by using the DMFT+ED method in conjunction with the linearized Eliashberg equation. 
All of the critical interactions towards the magnetic, orbital and superconducting instabilities have been found to be suppressed as compared to the RPA results. 
Remarkably, the $s_{++}$-pairing phase due to the orbital fluctuation is largely expanded as compared to the RPA result, while the $s_{\pm}$-pairing phase due to the magnetic fluctuation is reduced. 
This is caused by the local correlation effects which enhance the local, i. e., the ${\bm q}$-independent magnetic (orbital) fluctuation resulting in the local component of the repulsive (attractive) pairing interaction responsible for the suppression (enhancement) of the $s_{\pm}$ ($s_{++}$)-pairing. 
Although the case with $U<U'$ is not realistic and the FO fluctuation enhanced there ($d_{xy}$ intra-orbital component) is not corresponding to the softening of $C_{66}$, the same effects due to the local correlation are expected to occur in the $s_{++}$-pairing in the realistic cases with the electron-phonon interaction\cite{yanagi_4,yanagi_2,kontani} and/or the mode-coupling effects of the Coulomb interaction\cite{onari} and will be discussed in subsequent papers.

\section*{Acknowledgments}
This work was partially supported by a Grant-in-Aid for Scientific Research from the Ministry of Education, Culture,  Sports, Science  and Technology.

\bibliographystyle{jpsj.bst}
\bibliography{jpsj1.bib}
\end{document}